\documentclass[twocolumn,showpacs,aps,prb,superscriptaddress,floatfix,amsmath,amssymb]{revtex4}
\usepackage{graphicx}% Include figure files
\usepackage{dcolumn}% Align table columns on decimal point
\usepackage{bm}% bold math
\usepackage{subfigure}

\begin{document}
%\end{document}

\title{Revisiting the ground state of CoAl$_2$O$_4$: comparison to the conventional antiferromagnet MnAl$_2$O$_4$}

\author{G. J. MacDougall}
\email{gmacdoug@illinois.edu}
\affiliation{Department of Physics and Seitz Materials Research Laboratory, University of Illinois at Urbana-Champaign, Urbana, Illinois, 61801, USA}
\affiliation{Quantum Condensed Matter Division, Oak Ridge National Laboratory, Oak Ridge, Tennessee, 37831, USA}

\author{A. A. Aczel}
\affiliation{Quantum Condensed Matter Division, Oak Ridge National Laboratory, Oak Ridge, Tennessee, 37831, USA}

\author{Yixi Su}
\affiliation{J\"ulich Centre for Neutron Science JCNS at Heinz Maier-Leibnitz Zentrum (MLZ), Forschungszentrum J\"ulich GmbH, Lichtenbergstr. 1, D-85747 Garching, Germany}

\author{W. Schweika}
\affiliation{J\"ulich Centre for Neutron Science JCNS, Forschungszentrum J\"ulich GmbH, 52425 J\"ulich, Germany}

\author{E. Faulhaber}
\affiliation{J\"ulich Centre for Neutron Science JCNS at Heinz Maier-Leibnitz Zentrum (MLZ), Forschungszentrum J\"ulich GmbH, Lichtenbergstr. 1, D-85747 Garching, Germany}
\affiliation{Helmholtz-Zentrum Berlin f\"ur Materialien und Energie, Hahn-Meitner-Platz 1, D-14109 Berlin}

\author{A. Schneidewind}
\affiliation{Forschungsneutronenquell Heinz Meier-Leibnitz (FRM-II), D-85747 Garching, Germany}
\affiliation{Helmholtz-Zentrum Berlin f\"ur Materialien und Energie, Hahn-Meitner-Platz 1, D-14109 Berlin}

\author{A. D. Christianson}
\affiliation{Quantum Condensed Matter Division, Oak Ridge National Laboratory, Oak Ridge, Tennessee, 37831, USA}

\author{J. L. Zarestky}
\affiliation{Division of Materials Science and Engineering, Ames Laboratory, Iowa State University, Ames, Iowa 50011, USA}

\author{H. D. Zhou}
\affiliation{Department of Physics and Astronomy, University of Tennessee, Knoxville, Tennessee 37996, USA}
\affiliation{National High Magnetic Field Laboratory, Tallahassee, Florida 32310, USA}

\author{D. Mandrus}
\affiliation{Department of Materials Science and Engineering, University of Tennessee, Knoxville, Tennessee 37996, USA}
\affiliation{Materials Science and Technology Division, Oak Ridge National Laboratory, Oak Ridge, Tennessee, 37831 USA}

\author{S. E. Nagler}
\affiliation{Quantum Condensed Matter Division, Oak Ridge National Laboratory, Oak Ridge, Tennessee, 37831, USA}
\affiliation{Bredesen Center, University of Tennessee, Knoxville, TN, 37996, USA}

\date{\today}

\begin{abstract}
The A-site spinel material, $\mathrm{CoAl_2O_4}$, is a physical realization of the frustrated diamond-lattice antiferromagnet, a model in which unique incommensurate or `spin-spiral liquid' ground states are predicted. Our previous single-crystal neutron scattering study instead classified it as a `kinetically-inhibited' antiferromagnet, where the long ranged correlations of a collinear N$\mathrm{\acute{e}}$el ground state are blocked by the freezing of domain wall motion below a first-order phase transition at T* = 6.5 K. The current paper provides new data sets from a number of experiments, which support and expand this work in several important ways. We show that the phenomenology leading to the kinetically-inhibited order is unaffected by sample measured and instrument resolution, while new low temperature measurements reveal spin correlations are unchanging between T = 2 K and 250 mK, consistent with a frozen state. Polarized diffuse neutron measurements show several interesting magnetic features, which can be entirely explained by the existence of short-ranged N$\mathrm{\acute{e}}$el order. Finally, and crucially, this paper presents some of the first neutron scattering studies of single crystalline $\mathrm{MnAl_2O_4}$, which acts as an unfrustrated analogue to $\mathrm{CoAl_2O_4}$ and shows all the hallmarks of a classical antiferromagnet with a continuous phase transition to N$\mathrm{\acute{e}}$el order at $T_N$ = 39 K. Direct comparison between the two compounds indicates that $\mathrm{CoAl_2O_4}$ is unique, not in the nature of high-temperature diffuse correlations, but rather in the nature of the frozen state below $T^*$. The higher level of cation inversion in the $\mathrm{MnAl_2O_4}$ sample indicates that this novel behavior is primarily an effect of greater next-nearest-neighbor exchange.
\end{abstract}

%PACS numbers {spin arrangements using neutrons, spinwaves, spin glasses and random magnets, Neutron inelastic}

\maketitle

The A-site spinels, $\mathrm{AB_2X_4}$ with $A$ magnetic, have seen a surge of interest in the past decade, due to a series of interesting experimental observations\cite{fritsch04,krimmel05,tristan05,buttgen06,krimmel06,krimmel06_2,mucksch07,suzuki07,krimmel09,zaharko10,macdougall11,zaharko11,roy13,zaharko14,nair14} and theoretical predictions of novel spin-liquid ground states\cite{bergman07,lee08,bernier08,chen09,chen09_2}. Magnetic cations in these materials comprise a bi-partite diamond lattice, and novel behavior is argued to be the result of a competition between nearest ($J_1$) and next-nearest ($J_2$) neighbor superexchange interactions\cite{roth64,fritsch04}. This is demonstrated explicitly by the calculations of Bergman \textit{et al.}\cite{bergman07}, who have shown for spin-only materials that the collinear N$\mathrm{\acute{e}}$el structure favored by $J_1$ is progressively destabilized with increasing $J_2$, until a Lifshitz point is encountered at $\frac{J_2}{J_1} = \frac{1}{8}$. For greater $J_2$, the ground state is predicted to be a novel `spiral spin liquid' (SSL) state characterized by fluctuations between an infinitely degenerate set of incommensurate spin spirals, whose propagation wavevectors form a series of calculable manifolds in reciprocal space (`spiral surfaces'). Further calculations\cite{bergman07,lee08,bernier08} predict that these mass degeneracies are lifted by low-lying thermal or quantum fluctuations, driving first-order phase transitions via the order-by-disorder mechanism\cite{villain80} to either unique spin spiral or N$\mathrm{\acute{e}}$el ordered states, depending on the specific value of $J_2$.

These ideas have been quite successful in explaining the multiple transitions and incommensurate ground state of the singular material $\mathrm{MnSc_2S_4}$\cite{krimmel06_2,mucksch07}, and very recent work has confirmed key SSL predictions\cite{gao16}. In oxides, however, the search for SSL physics has focussed most intently on the compound $\mathrm{CoAl_2O_4}$. In this material, the diamond sublattice of the spinel structure is occupied by Co$^{2+}$ cations, which exist in a tetrahedrally coordinated crystal field environment and have S=3/2. Early magnetization and heat capacity measurements by Tristan \textit{et al.} revealed a high degree of frustration and an `unconventional glass-like' transition at $T_g$ = 4.8 K\cite{tristan05}. Neutron powder diffraction (NPD) measurement by the same group revealed significant short-range antiferromagnetic correlations in the glassy state\cite{krimmel06}, and Bergman \textit{et al.} immediately listed $\mathrm{CoAl_2O_4}$ as a candidate SSL material\cite{bergman07}. Subsequent experimental work, however, has not settled the issue. Though Suzuki \text{et al.} concluded a spin liquid ground state with $T^* \sim$ 10 K based on results of thermodynamic measurements\cite{suzuki07}, a detailed NPD and NMR study of high-quality powders led Roy \textit{et al.} to declare the existence of a continuous phase transition to classical N$\mathrm{\acute{e}}$el state at $T_N$ = 9.8 K\cite{roy13}. Separate NPD\cite{zaharko10} and single crystal neutron\cite{zaharko11,zaharko14} studies by Zaharko $\textit{et al.}$ instead concluded an unconventional antiferromagnetic phase  below $T_N$ = 8 K, with a significant level of microstructural defects and accompanied by diffuse `streaks' of scattering which they suggested was indicative of co-existing SSL correlations.

Our own single-crystal neutron scattering study presented a yet different perspective on the low-temperature state of $\mathrm{CoAl_2O_4}$\cite{macdougall11}. As with other groups\cite{zaharko11,roy13}, we observed short-range antiferromagnetic correlations at moderate temperatures, with growing intensity and correlation length as the system was cooled toward an apparently `glass-like' transition at $T^*$ = 6.5 K. Contrary to expectations for a glass however, we showed that $T^*$ is associated with the sudden emergence of a second, anisotropic contribution to the scattering function with Lorentzian-squared character, and an inelastic spectrum dominated by collective spin-wave excitations. A Lorentzian-squared lineshape, we pointed out, is the expected form for scattering from a series of ordered domains separated by sharp walls, as seen during the coarsening step of order formation below a first-order phase transition\cite{nagler88}. Fits of measured spin-wave dispersions in our samples implied that $\frac{J_2}{J_1} \sim 0.104$- close to the Lifshitz point of $\frac{J_2}{J_1} = 0.125$ and near where Bergman \textit{et al.} predicted a first-order transition to a N$\mathrm{\acute{e}}$el ordered state\cite{bergman07}. In a frustrated spin system, the energy scale of exchange interactions are typically large compared to thermal energies at temperatures below observed magnetic ordering transitions, and one may well expect domain wall motion below any first-order transition to be `kinetically inhibited', leading to a frozen short-range ordered state such as the one we have reported\cite{macdougall11}. Fundamentally out of equilibrium, such a frozen coarsening picture would also naturally explain the glass-like, but unconventional, characteristics reported for bulk thermodynamic data\cite{tristan05,suzuki07}.

Nonetheless, the presence of site defects undoubtedly plays a role in determining the material properties of A-site spinels. Calculations by Savary \textit{et al.} suggest that spiral order in A-site spinels may be induced by relatively small numbers of point defects and that materials close to the predicted Lifshitz point may be predisposed to glassiness, at least when $\frac{J_2}{J_1} > 1/8$\cite{savary11}. Monte Carlo simulations by Zaharko \textit{et al.} for systems with $\frac{J_2}{J_1}=0.1$ predict that N$\mathrm{\acute{e}}$el order is weakened and diffuse scattering intensified when either systematic absences are introduced on the spinel A-site or impurity moments placed on the B-site\cite{zaharko14}. In the latter case, the presence of as few as 4$\%$ impurity moments was sufficient to suppress N$\mathrm{\acute{e}}$el order completely\cite{zaharko14}. Hanashima \textit{et al.} highlighted the effects of common cation site-inversion through a systematic magnetization study of the series $\mathrm{(Co_{1-\delta}Al_\delta)(Al_{2-\delta}Co_\delta)O_4}$. They suggest that either a spin glass phase with an unvarying $T_g$ = 4.5 K or a spin liquid ground state with $T^* \sim$ 7-10 K is favored for $\delta > 0.101$ and $\delta < 0.0643$, respectively, with a sizeable coexistence region at intermediate inversion levels\cite{hanashima13}. It should be kept in mind, however, that experimentally measured power laws for temperature dependence of heat capacity\cite{tristan05,suzuki07,macdougall11,roy13}  and frequency dependence of ac-susceptibility\cite{hanashima13} have consistently shown that $\mathrm{CoAl_2O_4}$ cannot be described as a canonical spin glass; it is not clear if and how conjectured `unconventional spin glass' states with strong antiferromagnetic correlations might be distinct from reports of collinear antiferromagnetic order with broader than resolution-limited Bragg peaks\cite{roy13}, `unconventional' N$\mathrm{\acute{e}}$el order with significant `microstructural effects'\cite{zaharko11}, or `kinetically inhibited order' characterized by finite-size ordered domains\cite{macdougall11}.

In the current article, we seek to clarify the nature of magnetic correlations in $\mathrm{CoAl_2O_4}$ and the greater aluminate family by presenting two complementary sets of experiments. In the first, we provide what is to our knowledge the first comprehensive single-crystal neutron scattering study of the material $\mathrm{MnAl_2O_4}$, which acts as an unfrustrated analogue to $\mathrm{CoAl_2O_4}$. The two materials have near identical structures, but the larger S=5/2 moments and weaker next-neighbor interactions in $\mathrm{MnAl_2O_4}$ reduce frustration and result in a continuous ordering transition at $T_N \sim$ 40 K to a collinear N$\mathrm{\acute{e}}$el ordered ground state\cite{roth64,tristan05,krimmel09}. Though recent reports of remnant paramagnetism\cite{tristan05} and short range magnetic correlations\cite{nair14} in the ordered state have led to speculations about competing states in this compound, our results seem to demonstrate that it acts as a model classical antiferromagnet with resolution-limited Bragg peaks and collective spin-wave excitations. Fits of measured dispersions imply $\frac{J_2}{J_1} = 0.069 \pm 0.006$, which is consistent with the expectations of Bergman \text{et al.}\cite{bergman07} for the known $T_N$. Our measurements were also able to distinguish the critical correlations in this material, which were seen to have a Lorentzian-like character and give rise to distinctive diffuse scattering to temperatures well above $T_N$, but still clearly associated with the N$\mathrm{\acute{e}}$el ordered state.

Our second set of experiments might be thought of as a confirmation and expansion of our original study on single-crystalline $\mathrm{CoAl_2O_4}$\cite{macdougall11}, with which we provide several new insights about this material. Whereas our previous study demonstrated Lorentzian-like N$\mathrm{\acute{e}}$el correlations in $\mathrm{CoAl_2O_4}$ suddenly giving way to anisotropic Lorentzian-squared correlations at transition temperature $T^*$ = 6.5K, the current manuscript shows with new data that these experimental results depend on neither the exact sample measured nor on the instrument resolution, and further that neither scattering lineshape nor correlation length changes with temperature below T = 2 K down to T = 250 mK. This lack of low-temperature evolution is particularly striking, as it seems to favor a frozen ground state picture over one which invokes a role for low-lying thermal excitations. New inelastic measurements again confirm that the inelastic spectrum is dominated by collective spin-wave excitations consistent with N$\mathrm{\acute{e}}$el order, but improved energy resolution in our measurements allow us to extract a more precise value for $\frac{J_2}{J_1} = 0.110 \pm 0.003$- even closer to the first-order transition region predicted by Bergman \textit{et al.}  Finally, we present new polarized neutron diffraction measurements of the significant magnetic diffuse scattering in this material, and show it to be isotropic with no measurable chirality. Simple simulations demonstrate that diffuse features are largely accounted for by the Lorentzian-like N$\mathrm{\acute{e}}$el correlations mentioned above.

Perhaps most crucial of all, direct comparison of results on $\mathrm{CoAl_2O_4}$ and $\mathrm{MnAl_2O_4}$ make clear the origin of the Lorentzian scattering in both compounds is simply critical correlations, and show that the cobalt material is unique, not in the nature of diffuse correlations at intermediate temperatures, but in the unusual character of the frozen state below $T^*$. As the $\mathrm{MnAl_2O_4}$ has the greater level of cation inversion, our results further reinforce the primary role of $\frac{J_2}{J_1}$ and not disorder in determining the magnetic ground state in these materials.

\section{Crystal growth and characterization}

\begin{figure}[tb]
\begin{center}
\includegraphics[width=\columnwidth]{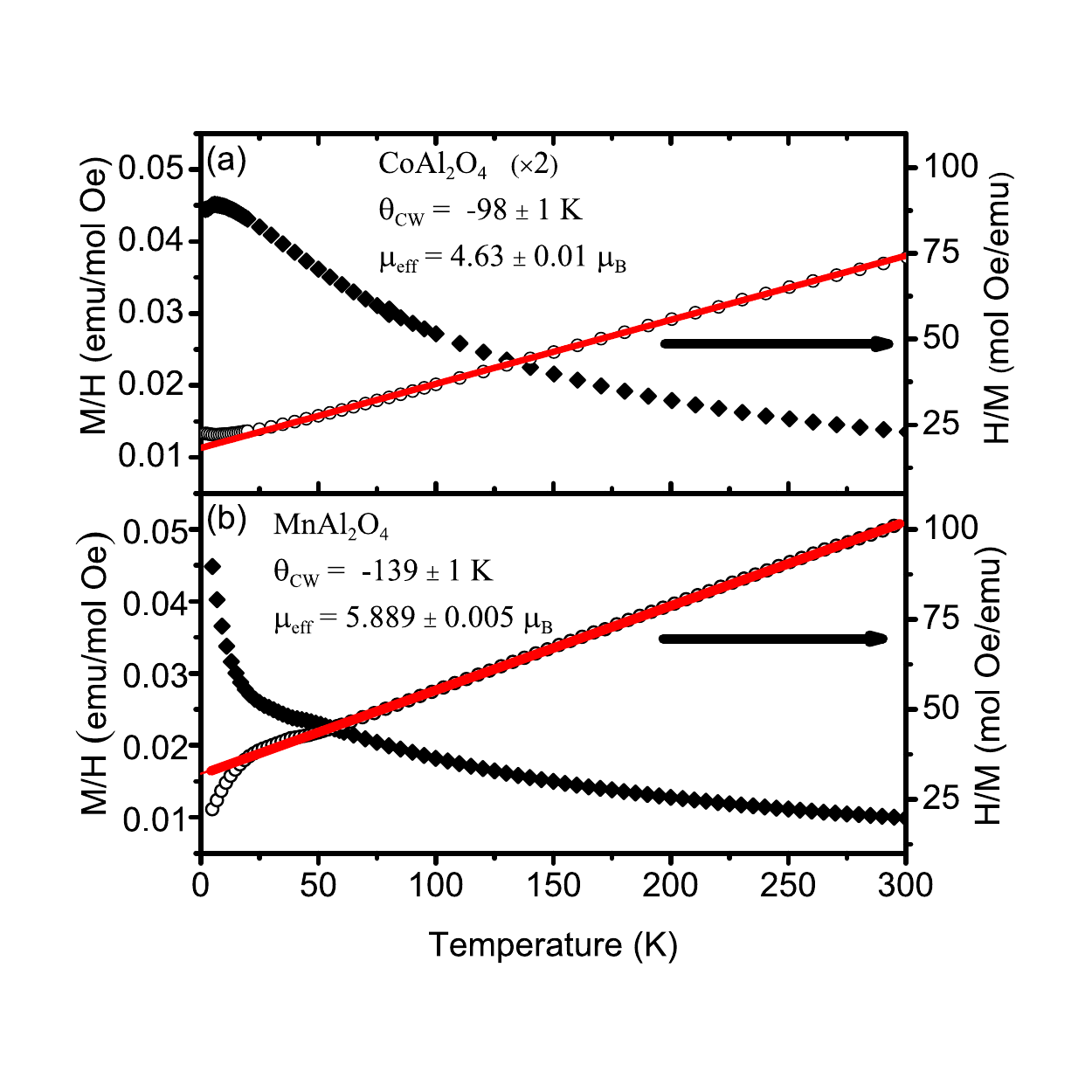}
\caption{Magnetization (closed symbols) of (a) $\mathrm{CoAl_2O_4}$ and (b) $\mathrm{MnAl_2O_4}$ single crystals, scaled by the applied fields of H = 10 kOe and 2 kOe, respectively. Also shown is the inverse susceptibility (open symbols) of each, and the lines of best fit of high temperature data to a linear dependence. Data for $\mathrm{CoAl_2O_4}$ has been scaled by a factor of two to facilitate comparison with $\mathrm{MnAl_2O_4}$ on the same scale. Effective moments and Weiss parameters listed were extracted from inverse fits.}\label{fig:SQUID}
\end{center}
\end{figure}

The single crystal samples explored in this study were grown using 2-mirror NEC optical image furnaces, at the National High Magnetic Field Laboratory (NHMFL) in Tallahassee, FL  and at Oak Ridge National Laboratory (ORNL), in Oak Ridge, TN. X-ray powder diffraction (XPD) was performed on crushed portions of the crystals to ensure sample purity and to quantify the level of cation site inversion. Bulk magnetization was subsequently measured using Quantum Design MPMS SQUID magnetometers, with main results shown in Fig.~\ref{fig:SQUID}.

The $\mathrm{CoAl_{2}O_{4}}$ crystals were grown at ORNL, and representative characterization data can be found in the Supplementary Information for Ref.~\onlinecite{macdougall11}. As reported previously, XRD refinements suggest a cation inversion $x = 0.02 \pm 0.04$, among the lowest in the literature. Our previous fits of magnetization to a Curie-Weiss temperature dependence also yielded an effective moment $\mu_{eff} = 4.89 \pm 0.03 \mu_B$ and Weiss constant $\Theta = -109 \pm 1$ K. The large effective moment for Co$^{2+}$ in a tetrahedral crystal field environment can be understood as an effect of spin-orbit coupling\cite{schlapp32,cossee60}, but it has recently been pointed out that this value is also larger than many other reports in the literature\cite{roy13}. On this last issue, we note that these other publications typically extract moments and Weiss constants from fits to inverse susceptibility, rather than magnetization itself with a background term and used a different temperature range. For more direct comparison to these studies, we reproduce the data from Ref.~\onlinecite{macdougall11} in Fig.~\ref{fig:SQUID}(a) and fit the inverse susceptibility over the range 100 K $< T <$ 300 K, neglecting the possible existence of temperature-independent background contributions. These fits yield the lower values $\mu_{eff} = 4.63 \pm 0.01 \mu_B$ and $\Theta = -98 \pm$ 1 K. Although these results put our samples more in line with other reports in the literature\cite{roy13}, the strong dependence of fit parameters on details of fitting procedure should preclude one from drawing strong conclusions from such an analysis. A more reliable aspect of the data in Fig.~\ref{fig:SQUID}(a) is the sharp cusp at $T^* = 6.5\pm 1$ K, which is above the universal glass temperature reported by Hanashima for materials with larger $x$\cite{hanashima13}, and which we have shown from neutron diffraction is consistent with a kinetically inhibited first-order phase transition\cite{macdougall11}.

The $\mathrm{MnAl_{2}O_{4}}$ crystal was grown and characterized at the NHMFL, and the inversion parameter was determined from XRD to be $x = 0.05 \pm 0.02$. This value is equal or higher than in the cobalt sample. Magnetization measurements were performed using an applied field H = 2 kOe in a field-cooled (FC) configuration, and the resultant data are shown in Fig.~\ref{fig:SQUID}(b). As with $\mathrm{CoAl_{2}O_{4}}$, high temperature data are well described by a Curie-Weiss temperature dependence, and were fit in an analogous way. The best fit yielded parameters $\mu_{eff} = 5.89 \pm 0.01 \mu_B$ and $\Theta = -139 \pm 1$ K. These values are comparable to the results of Tristan \textit{et al}\cite{tristan05}. A local maximum in susceptibility reveals an ordering temperature of $T_N \sim 40$ K, consistent with neutron results\cite{krimmel06}. Below $T_N$, there is a significant level of remnant paramagnetism, and a similar fitting procedure yields $\mu'_{eff} = 2.8 \pm 0.1 \mu_B$ and $\Theta' = -17 \pm 1$ K. These results are also in line with previous results\cite{tristan05}.

\section{Neutron scattering}

\subsection{Experimental methods}

Neutron scattering measurements presented in this work were performed using instruments at the High Flux Isotope Reactor (HFIR) at ORNL and the Heinz Maier-Leibnitz Zentrum (MLZ) at Garching, Germany. $\mathrm{MnAl_{2}O_{4}}$ was explored using the HB1 spectrometer at HFIR, with a 3g single crystal mounted in the (H H L) scattering plane, and collimation set to  48$^{\prime}$-40$^{\prime}$-40$^{\prime}$-120$^{\prime}$. Elastic and inelastic measurements were performed with fixed final energies of $E_f$ = 13.5 meV and 14.7 meV, respectively.

New elastic measurements of $\mathrm{CoAl_{2}O_{4}}$ were performed first using the HB1a triple-axis  instrument with $E_i$ = 14.6 meV neutrons and collimation 48$^{\prime}$-48$^{\prime}$-40$^{\prime}$-68$^{\prime}$, which are identical to conditions employed in Ref.~\onlinecite{macdougall11}. One crystal was wired to a copper mounting plate in the (H H L) scattering plane, and mounted to the cold finger of a He3 cryostat insert to achieve lowest temperatures. Subsequent measurements used 48$^{\prime}$-20$^{\prime}$-20$^{\prime}$-68$^{\prime}$ collimination, which reduced the resolution width by 25$\%$ in \textbf{Q} and 11$\%$ in energy. Both HB1a and HB1 measurements employed pyrolytic graphite (PG) monochromators and analyzers, and further PG filters were used to remove higher order contamination in the neutron beam. Inelastic measurements were performed using the PANDA\cite{schneidewind15} cold triple-axis spectrometer at MLZ, using a configuration with a double-focussing analyzer and monochromator and fixed $E_f$ = 5 meV. Higher order contamination was removed using a cooled beryllium filter. Also at MLZ, diffuse scattering measurements were carried out using the polarized neutron spectrometer, DNS\cite{schweika01,su15}, with a neutron wavelength of 4.75 \AA.

\subsection{$\mathrm{MnAl_{2}O_{4}}$}

\begin{figure}[t]
\begin{center}
\includegraphics[width=\columnwidth]{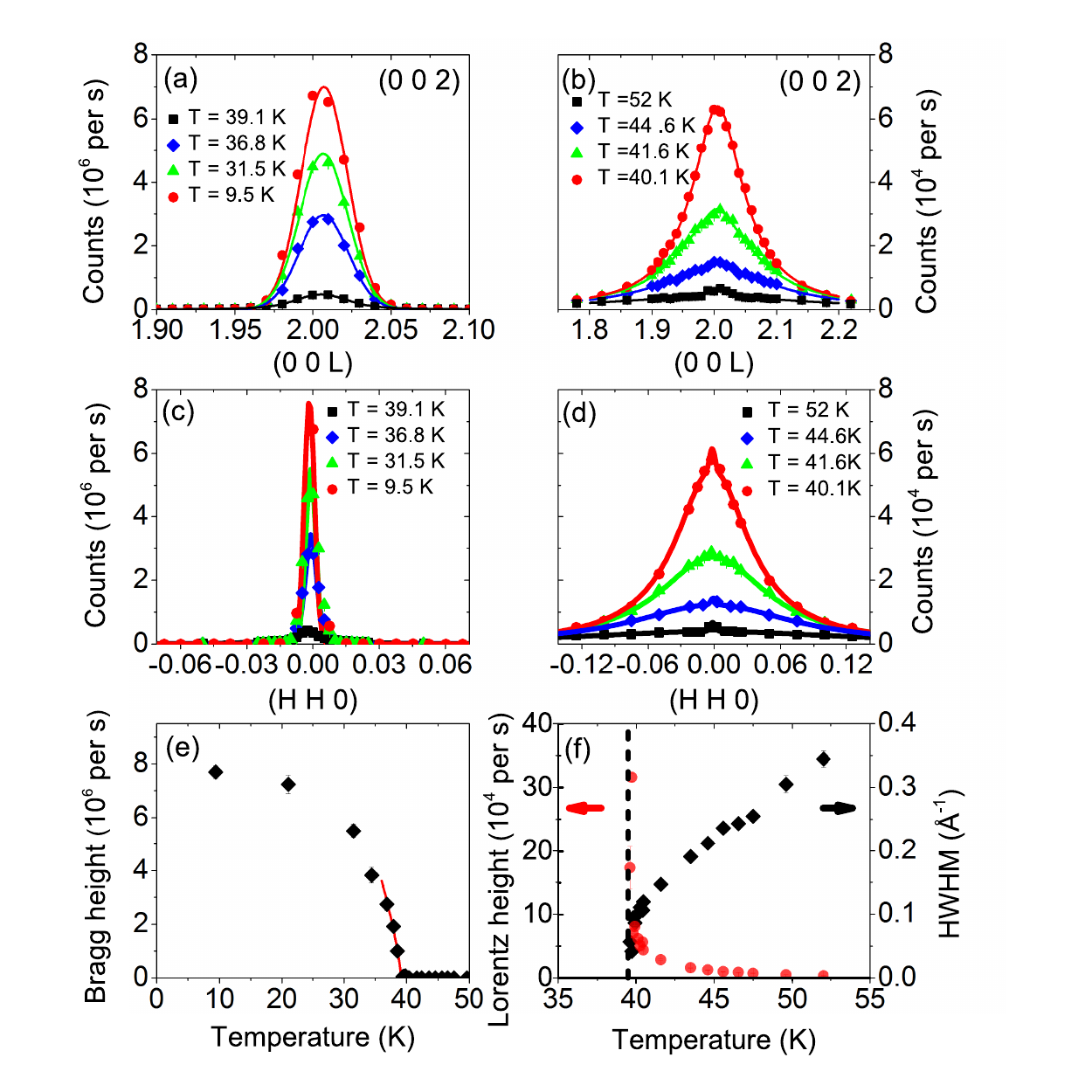}
\caption{Elastic neutron scattering measurements of magnetic correlations in $\mathrm{MnAl_2O_4}$. \textbf{(a)/(b):} Representative plots of neutron scattering intensity as \textbf{Q} is scanned across the magnetic (0 0 2) Bragg peak in the [0 0 L] direction. Panel (a) ((b)) contains scans at temperatures below (above) the N$\mathrm{\acute{e}}$el temperature, $T_N$, determined below. \textbf{(c)/(d):} Equivalent scans in the [H H 0] direction.  Solid lines in these plots represent fits to Gaussian + Lorentzian lineshapes, as described in the main text. \textbf{(e):} Temperature dependence of the Gaussian intensity, associated with magnetic long-range order. The solid line represents a power law fit used to determine $T_N$. \textbf{(f)} Temperature dependences of the Lorentzian height and width, associated with magnetic critical fluctuations. The dashed line denotes the position of $T_N$, determined in (e). }\label{fig:MnAl2O4_elastic}
\end{center}
\end{figure}

We begin by presenting the results of our neutron scattering study of a single-crystal of $\mathrm{MnAl_2O_4}$, in which NPD measurements have long implied the existence of collinear N$\mathrm{\acute{e}}$el order below $T_N \sim$ 40 K\cite{roth64,krimmel06}. Our measurements confirm the same ordered state in crystals. Significant magnetic scattering intensity is seen at T = 8 K above measured intensity at T = 50 K at the positions of (H K L) = (0 0 2), (1 1 1), (1 1 3), (1 1 5), (2 2 2) and (3 3 3) Bragg peaks, but not at (0 0 4) or (2 2 0), in line with expectations for the N$\mathrm{\acute{e}}$el state. Here, (H K L) are the reciprocal lattice indices of the cubic unit cell. To track the evolution of magnetic correlations with temperature, we performed a series of radial and transverse scans across (0 0 2) and (1 1 1) magnetic Bragg positions, with the main results for the (0 0 2) peak illustrated in Fig.~\ref{fig:MnAl2O4_elastic}.

Plots of representative radial (Fig.~\ref{fig:MnAl2O4_elastic} (a) and (b)) and transverse (Fig.~\ref{fig:MnAl2O4_elastic} (c) and (d)) scans across the (0 0 2) position demonstrate that the N$\mathrm{\acute{e}}$el correlations in this compound are nearly isotropic at all temperatures, and consist of two distinct forms of scattering. Intense, resolution-limited Gaussian peaks dominate the scattering at temperatures below $T_N$ (Figs.~\ref{fig:MnAl2O4_elastic} (a) and (c)), and a two orders-of-magnitude weaker Lorentzian component is also present, which is clearest at temperatures directly above $T_N$ (Figs.~\ref{fig:MnAl2O4_elastic} (b) and (d)). Radial and transverse scans across the (0 0 2) peak were fit simultaneously to a sum of a Gaussian and a Lorentzian convolved with instrument resolution, and fit parameters are shown in Figs.~\ref{fig:MnAl2O4_elastic} (e) and (f). Though not presented here, inspection of scattering at the (1 1 1) peak position leads to similar plots.

The temperature dependence of the Gaussian component clearly confirms that it acts as an order parameter for the collinear antiferromagnetic phase. Fits near the transition to a mean-field power law yield  $T_N$ = 39.1 $\pm$ 0.4 K, in-line with the value gleaned from magnetization measurements. The ordered moment size extracted from the integrated Bragg peak intensity at lowest temperature was $m_{Mn} = 5.2 \pm 0.2 \mu_B$, higher than originally reported by Roth\cite{roth64} but within error equal to the full moment expected for S = 5/2 spins. The weak Gaussian contribution to the scattering at highest temperatures is consistent with scattering from the (0 0 4) nuclear Bragg peak by $\lambda$/2 neutrons, imperfectly filtered from the incident beam. The Lorentzian component is evident at all temperatures, but masked by the much stronger Gaussian component below $T_N$. At higher temperatures, it is the dominant contribution to the scattering, and the peak height and correlation length (inverse of width) diverge with power law dependences as the temperature approaches $T_N$ from above. This is the exact behavior expected for scattering above a second order phase transition, and clearly associates the Lorentzian component with classical Ornstein-Zernicke critical correlations\cite{collins_book}.

\begin{figure}[t]
\begin{center}
\includegraphics[width=0.90\columnwidth]{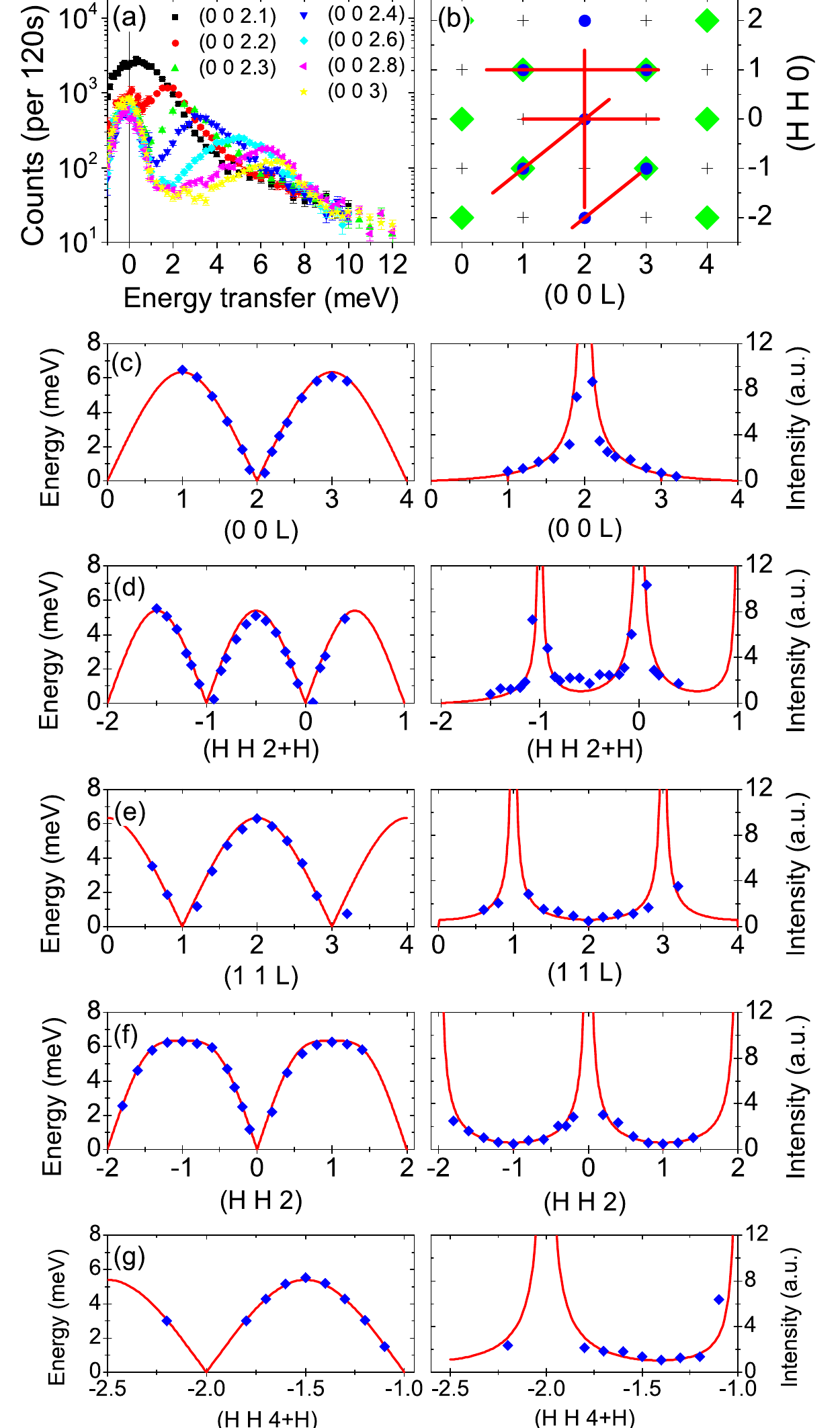}
\caption{Measurements of collective spin-wave excitations in $\mathrm{MnAl_2O_4}$. \textbf{(a):} A series of representative constant-\textbf{Q} scans at different points along the line [0 0 L] in reciprocal space, presented on a logarithmic scale. \textbf{(b):} Schematic of the [H H L] scattering plane, where diamonds (circles) represent the positions of nuclear (magnetic) Bragg peaks for a material with the spinel structure and N$\mathrm{\acute{e}}$el antiferromagnetism. Solid red lines represent the loci of paths along which the spin-wave dispersion was measured. \textbf{(c)-(g):} Plots of energy (left) and integrated intensity (right) of spin-wave excitations observed via constant-\textbf{Q} measurements. Solid lines represent the predictions of semiclassical spin-wave theory, using exchange parameters determined by fits of dispersion data only. Details of the fitting model are given in the main text.}\label{fig:MnAl2O4_SW}
\end{center}
\end{figure}

At low temperatures, inelastic neutron scattering measurements reveal the emergence of collective spin-wave excitations out of the N$\mathrm{\acute{e}}$el ordered state. Fig.~\ref{fig:MnAl2O4_SW} (a) shows a comparison of several energy scans at reciprocal lattice positions along the line [0 0 L], and demonstrates a variation of the excitation energy from E = 0 to 6 meV as reciprocal lattice vector \textbf{Q} is varied from the magnetic zone center (0 0 2) to the zone boundary (0 0 3). There was no indication of an excitation gap in $\mathrm{MnAl_{2}O_{4}}$ within the sensitivity of our measurements. Using similar constant-\textbf{Q} scans, we tracked the spin-wave excitations along five different paths in reciprocal space, indicated by solid red lines in Fig.~\ref{fig:MnAl2O4_SW} (b), and fit the results to predictions of classical spin-wave theory broadened to account for instrument resolution effects. Results are plotted in Figs.~\ref{fig:MnAl2O4_SW} (c)-(g). Spin-wave calculations assumed the  Hamiltonian

\begin{equation}
 \mathcal{H} = J_1 \sum\limits_{< i,j >}\mathbf{S_i} \cdot \mathbf{S_j} + J_2 \sum\limits_{<< i,j >>}\mathbf{S_i} \cdot \mathbf{S_j},
\end{equation}

where the first (second) sum is over nearest (next-nearest) neighbors on the Mn$^{2+}$ sublattice. Simultaneous fits of all five dispersion curves led to the determination that $J_1 = 0.87 \pm 0.02~ meV$ and $J_2 = 0.059 \pm 0.004~ meV$. These give the ratio, $\frac{J_2}{J_1} = 0.069 \pm 0.006$, which reproduce the measured ordering temperature within the predictions of Bergman \textit{et al.}\cite{bergman07}. Intensities were not fit, and lines in the five plots on the right side of Figs.~\ref{fig:MnAl2O4_SW} (c)-(g) were made using only a single free scaling parameter, after assuming the above determined values for $J_1$ and $J_2$.

% J1 = -0.4347 pm 0.0109

%J2 = -0.0295 pm 0.0020

%J2/J1= 0.06786 pm 0.0063

\subsection{$\mathrm{CoAl_{2}O_{4}}$}

The data presented in Figs.~\ref{fig:MnAl2O4_elastic} and \ref{fig:MnAl2O4_SW} for $\mathrm{MnAl_2O_4}$ immediately beg comparison to the existing neutron scattering results for single-crystalline $\mathrm{CoAl_2O_4}$\cite{macdougall11}. Much like $\mathrm{MnAl_2O_4}$, we previously observed that isotropic Lorentzian-like scattering emerges in $\mathrm{CoAl_2O_4}$ at moderate temperatures, with a growing intensity and narrowing peak width as the temperature $T^*\sim 6.5$ K is approached from above. Below this temperature, we reported a qualitative change in lineshape, consistent with the emergence of a second contribution to the scattering, though it had an anisotropic Lorentzian-squared character, in contrast to the Gaussian seen in Fig.~\ref{fig:MnAl2O4_elastic}. It is most natural then to interpret the Lorentzian component in $\mathrm{CoAl_2O_4}$, as well, as a signature of critical fluctuations. The Lorentzian-squared component should then be associated with some form of order and, as we pointed out above, is consistent with a kinetically inhibited state below a first-order phase transition. Within this interpretation, the persistence of intense Lorentzian scattering at lowest temperatures is a signature of remnant fluctuations below $T^*$, due to the incomplete nature of the transition.

\begin{figure}[ht]
\begin{center}
\includegraphics[width=\columnwidth]{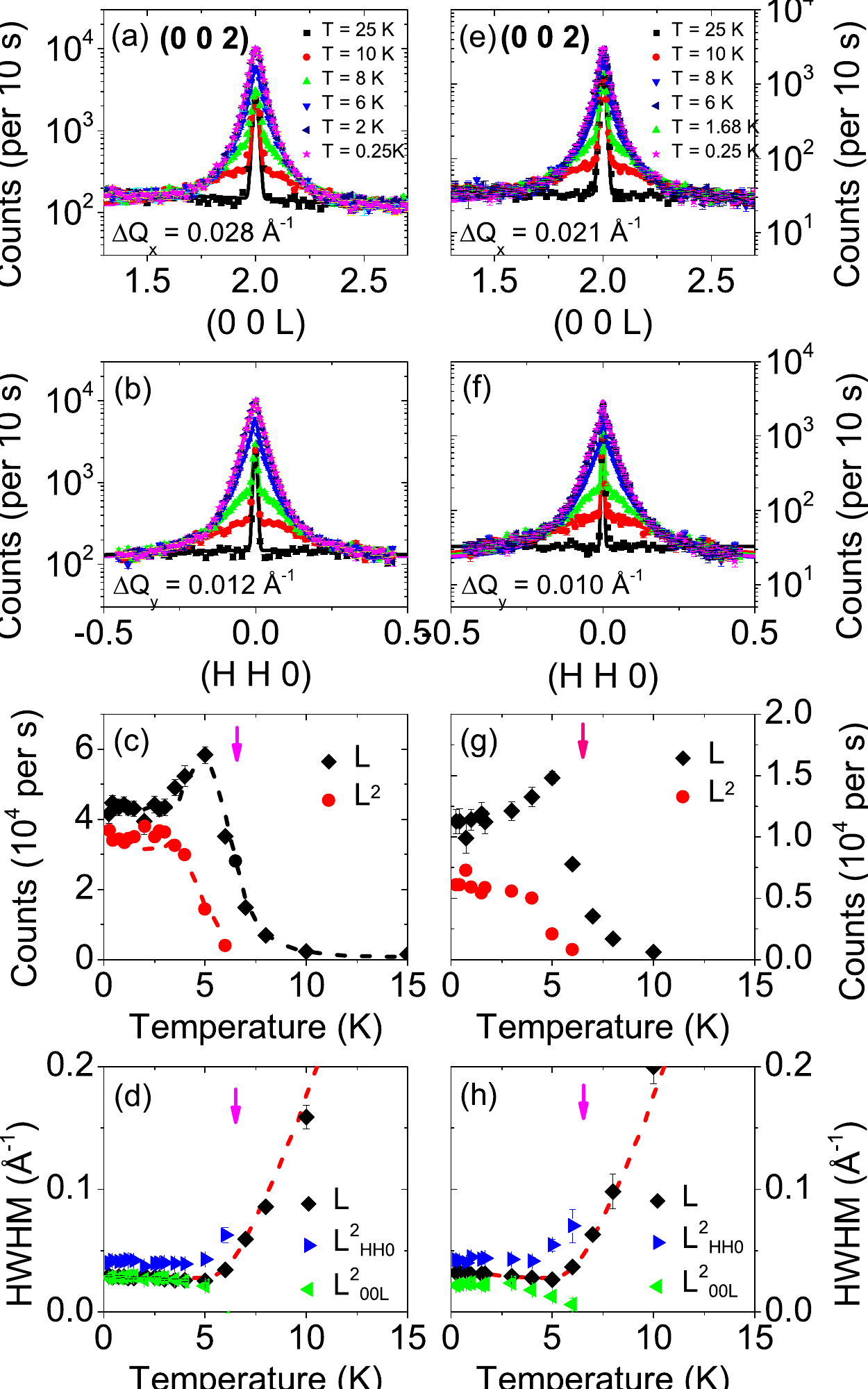}
\caption{Thermal triple-axis elastic measurements of magnetic correlations in $\mathrm{CoAl_2O_4}$. \textbf{(a)/(b):} Representative plots of neutron scattering intensity as \textbf{Q} is scanned across the magnetic (0 0 2) Bragg peak in the [0 0 L] (a) and [H H 0] (b) directions. Solid lines represent simultaneous fits to an isotropic Lorentzian plus anisotropic Lorentzian-squared fit function. \textbf{(c)/(d):} Heights (c) and widths (d) of fitted Lorentzian and Lorentzian-squared components. For direct comparison, dashed lines represent data from Ref.~\onlinecite{macdougall11}, taken on a different crystal with identical instrument resolution conditions. \textbf{(e)-(h):} Equivalent data presented in (a)-(d), but taken with the instrument in a higher resolution configuration. Specifics are given in main text. Vertical arrows in these plots denote the position of $T^*$, as determined by magnetization measurements.}\label{fig:HB1a}
\end{center}
\end{figure}

In Fig.~\ref{fig:HB1a}, we show data from new experiments which reproduce our original experimental results in a new single-crystal sample, and under more stringent experimental conditions. Using the same instrument and collimation choices as our previous work\cite{macdougall11}, we first measured elastic neutron scattering intensity in the vicinity of the magnetic (0 0 2) Bragg reflection, while crossing the peak in [0 0 L] and [H H 0] directions in reciprocal space and while decreasing temperature to T = 0.25 K. The resultant data are plotted in Figs.~\ref{fig:HB1a} (a) and (b), revealing the same evolution of Lorentzian to Lorentzian + Lorentzian-squared line-shape reported previously, and were fit the same way. Fit parameters are plotted in Figs.~\ref{fig:HB1a} (c) and (d), alongside data from our original measurements\cite{macdougall11} (dashed lines), and confirm the reproducibility of our results between different samples in the same temperature range. Figs.~\ref{fig:HB1a} (e)-(h) represent a second set of measurements performed using tighter instrument collimation (25$\%$ reduction in momentum resolution), and demonstrate the independence of our main conclusions from exact instrument resolution. Specifically, we have directly confirmed that the isotropic Lorentzian plus anisotropic Lorentzian-squared lineshape remains far superior to any single component scattering function, regardless of the instrument configuration.

Significantly, the data in Fig.~\ref{fig:HB1a} also reveal that both scattering lineshape and peak width remain unchanged in the temperature region T = 0.25 - 2 K. This is a direct contradiction to expectations from Monte Carlo calculations, which predict resolution limited (Gaussian) Bragg scattering at lowest temperatures and ascribe the finite scattering widths at T = 2 K to the presence of low-lying thermal excitations\cite{zaharko11}. In passing, we note that our low temperature results would also seem to rule out the existence of a second magnetic transition to a competing ordered state, as one might expect if the $T^* = 6.5$ K transition was driven by the order-by-disorder mechanism\cite{bergman07}. Temperature independence below 2K is perfectly in line with expectations for a glassy or kinetically frozen state, both of which are inherently out of thermal equilibrium. By considering the integrated weight under the (0 0 2) Bragg reflection, we estimate the frozen moment size to be $m_{Co} = 1.09 \pm 0.08 \mu_B$.

\begin{figure}[t]
\begin{center}
\includegraphics[width=\columnwidth]{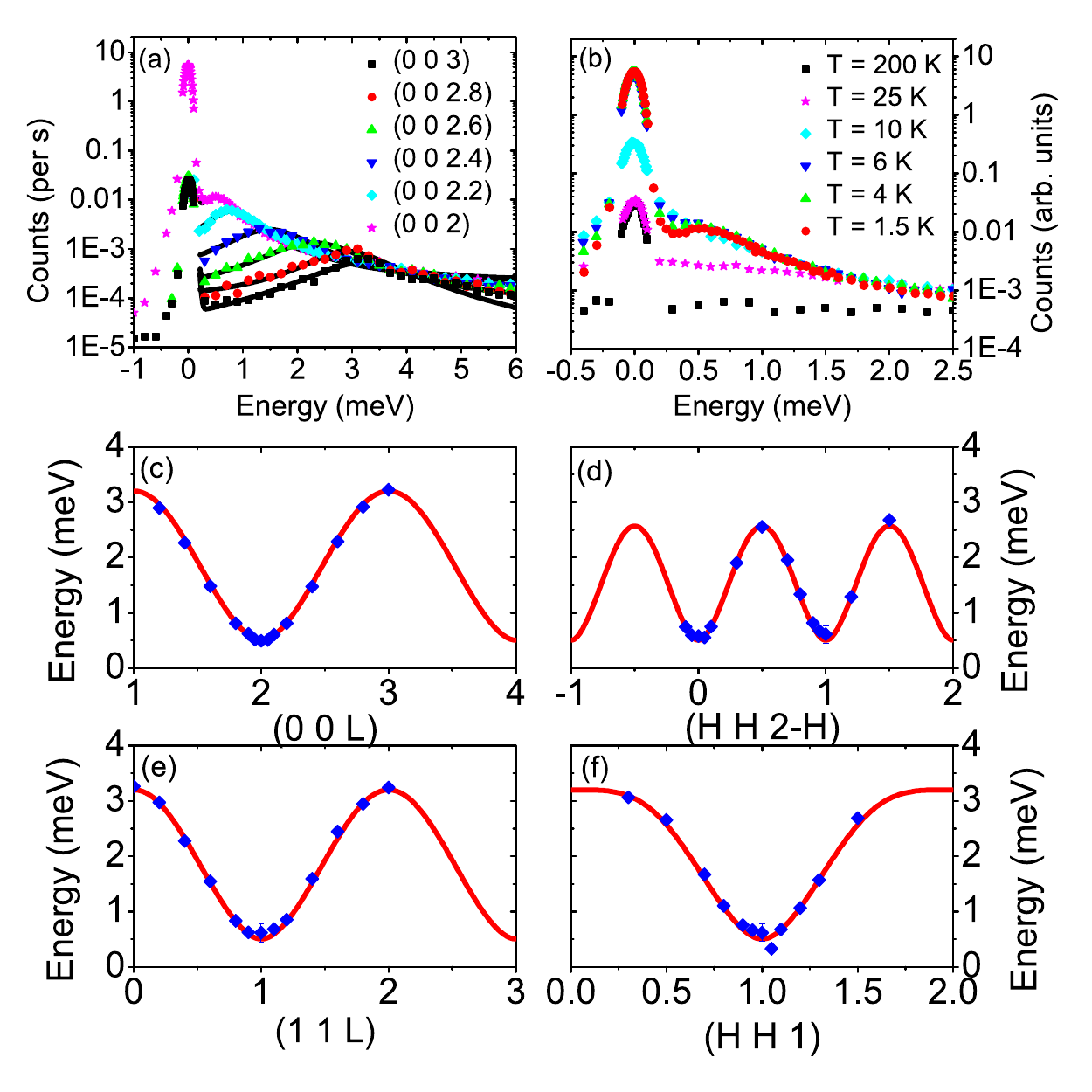}
\caption{Inelastic neutron scattering results for $\mathrm{CoAl_2O_4}$, from cold triple-axis measurements. \textbf{(a):} Representative scans of neutron scattering intensity versus energy at several reciprocal lattice positions along the [0 0 L] axis in reciprocal space. \textbf{(b):} Neutron scattering intensity versus energy at the [0 0 2] position, for several different temperatures.  \textbf{(c)-(f):} Plots showing the variation of collective spin-wave excitations along four different paths in reciprocal space. Solid lines in these plots represent best fits to the spin-wave model described in the main text.}\label{fig:PANDA}
\end{center}
\end{figure}

To search for mode splitting, novel excitations or other signatures of a competing ground state, we accompanied our high-resolution elastic measurements with new inelastic scattering measurements using a cold neutron triple-axis spectrometer, with a significant improvement in energy resolution ($\Delta E\sim$ 0.1 meV, vs 0.6 meV in previous experiments). As with previous measurements however, the resultant data, shown in Fig.~\ref{fig:PANDA}, failed to reveal any overt signature of order beyond the long-range collective spin-wave excitations associated with the N$\mathrm{\acute{e}}$el ordered state. Typical inelastic scans are shown in Fig.~\ref{fig:PANDA}(a), which demonstrates dispersion of the single gapped excitation along the line [0 0 L] in reciprocal space. Fig.~\ref{fig:PANDA}(b) shows the mode disappear as temperature is increased. Figs.~\ref{fig:PANDA}(b)-(f) show the variation of excitation energy with momentum transfer along five different paths in reciprocal space. As with previously reported data\cite{macdougall11,zaharko11}, these dispersions are perfectly captured by classical spin-wave theory, using the Hamiltonian

\begin{equation}
\mathcal{H} = J_1 \sum\limits_{< i,j >}\mathbf{S_i} \cdot \mathbf{S_j} + J_2 \sum\limits_{<< i,j >>}\mathbf{S_i} \cdot \mathbf{S_j} - g\mu_B H_A\sum\limits_{i}\|S_{z,i}\| ,
\end{equation}

where again the first (second) sum is over nearest (next-nearest) neighbors on the Co$^{2+}$ sublattice.

The best fit of the collective dispersions gives values $J_1 = 0.94 \pm 0.01 meV$, $J_2 = 0.103 \pm 0.002 meV$ and $g\mu_BH_A = 0.015 \pm 0.001 meV$ for nearest neighbor exchange, next-nearest neighbor exchange and anisotropy field, respectively. This gives a ratio of $\frac{J_2}{J_1} = 0.110 \pm 0.003$,  more precise but equal within error to our previous results\footnote{The factor of 2 difference between the current values of $J_1$ and $J_2$ and those reported in Ref.~\onlinecite{macdougall11} simply reflects the use a different convention for the summation in the Hamiltonian- over spins rather than bonds.}. The spin gap is observed to be $0.50 \pm 0.06 meV$, also consistent with our original work\cite{macdougall11}. We have previously argued that this gap is a signature of spin-orbit coupling on the cobalt site, but it is interesting to note that a gap on the order of $0.2 meV$ is predicted from finite-size effects if one considers the measured $\sim~$150\AA~ domain wall separation in the cubic [1 1 0] direction. Thus, the observed gap may be another signature of a frozen domain structure. Notably, the width of the excitation at the magnetic zone center ($\Delta E = 0.43\pm 0.08 meV$) also remains much broader than energy resolution; this indicates that the line broadening at the zone center is inherent, consistent with the failure of the material to achieve true long-range order.

\begin{figure*}[t]
\begin{center}
\includegraphics[width=2\columnwidth]{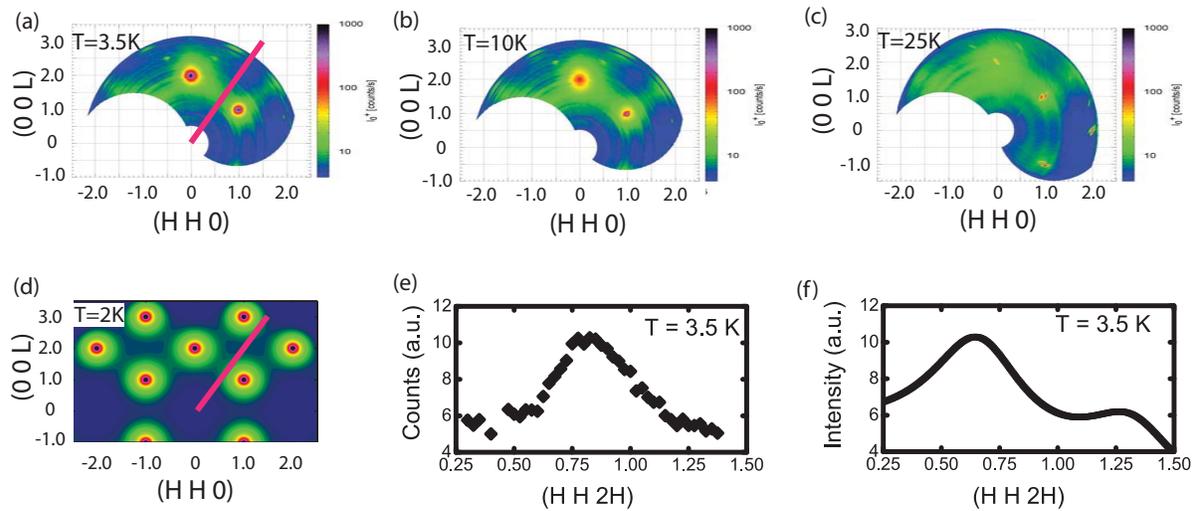}
\caption{Results of diffuse scattering measurements in the [H H L] plane of $\mathrm{CoAl_2O_4}$ crystals. \textbf{(a)-(c)} Plots of non-spin-flip scattering intensity in the out-of-plane (z) direction, which contains both magnetic and structural contributions. Included are data at temperatures T = 3.5 K (a), 10 K (b) and 25 K (c). \textbf{(d)} Simulation of the expected diffuse scattering at T = 3.5 K, assuming only the Lorentzian-like antiferromagnetic correlations inferred from triple-axis measurements. \textbf{(e)} Scattering intensity at T = 3.5 K as one travels from the origin along the [1 1 2] direction in reciprocal space (solid line in (a)).  \textbf{(f)} The expected scattering intensity along the same line, while assuming only contributions from the Lorentzian peaks simulated in (d). }\label{fig:DNS_sim_diffuse}
\end{center}
\end{figure*}

\begin{figure}[th]
\begin{center}
\includegraphics[width=1.1\columnwidth]{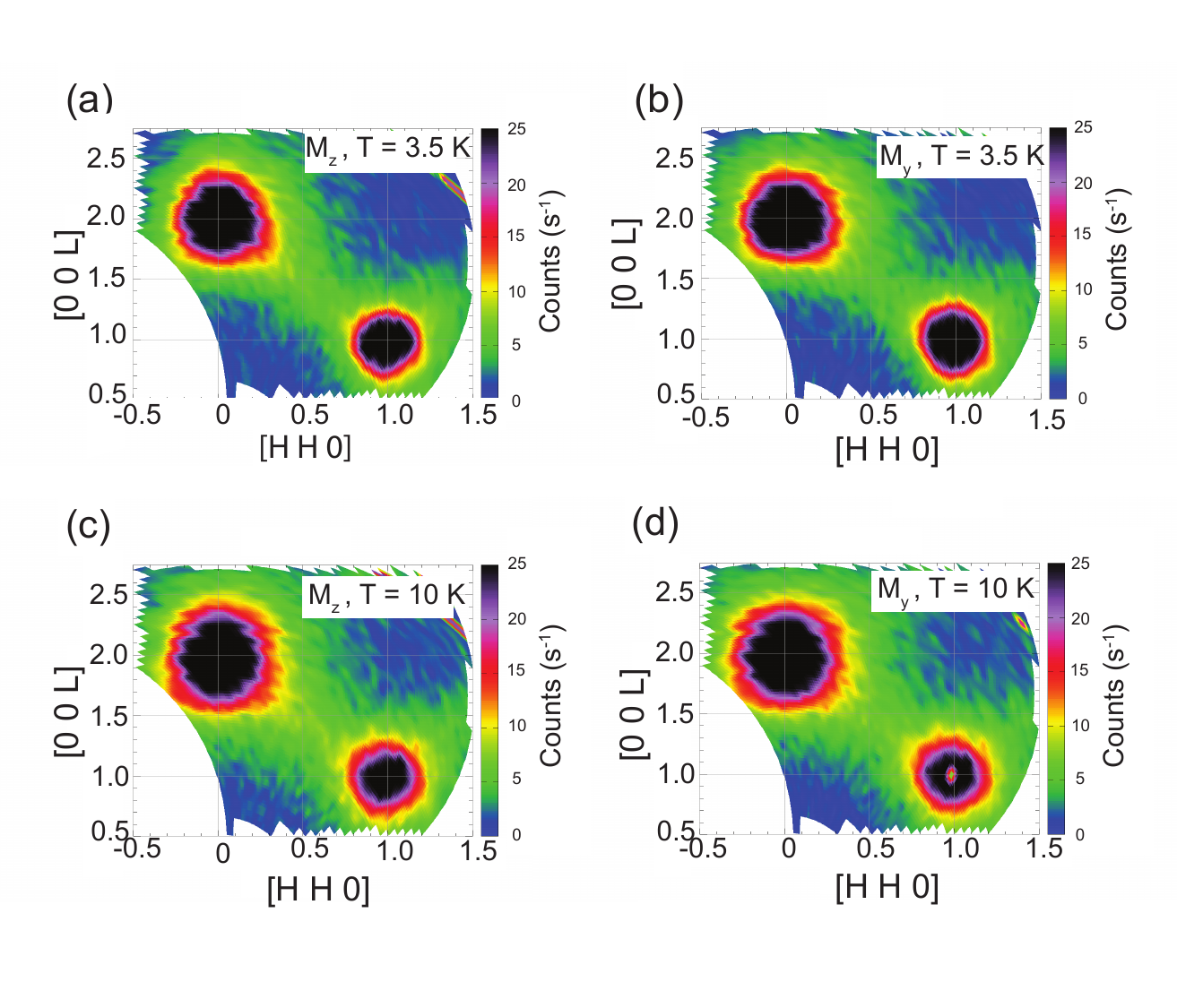}
\caption{Plots of the purely magnetic contribution to diffuse neutron scattering in $\mathrm{CoAl_2O_4}$, separated into in-plane (y) and out-of-plane (z) directions, and showing essentially isotropy of spin correlations. Panels (a)/(b) show results for measurements taken at T = 3.5 K and panels (c)/(d) at T = 10 K, below and above the transition temperature $T^*$ = 6.5 K.}\label{fig:DNS_mag_diffuse}
\end{center}
\end{figure}

Finally, to learn more about the nature of diffuse scattering in this material and search for signatures of SSL correlations, we performed a series of new polarized neutron diffraction measurements using the DNS instrument at MLZ. This instrument uses the XYZ method of polarization analysis, which allows one to extract information about diffuse magnetic correlations over a wide range of reciprocal space, including the directionality of ordered spins and possible existence of spin chirality\cite{schweika10}. Diffuse scattering correlations were explored in the [H H L] scattering plane of $\mathrm{CoAl_2O_4}$ with two sets of measurements, and main results are plotted in Fig.~\ref{fig:DNS_sim_diffuse} and Fig.~\ref{fig:DNS_mag_diffuse}.

Correlations were first explored over a 180$^{\circ}$ range of scattering angles using $\lambda$ = 4.75\AA~ neutrons and only investigating the polarization direction perpendicular to the scattering plane, which mixes magnetic and non-magnetic contributions. The results, shown in Fig.~\ref{fig:DNS_sim_diffuse} on a logarithmic scale, demonstrate that scattering in these materials is dominated by short-range antiferromagnetic correlations, in the form of Lorentzian-like peaks centered at Bragg positions associated with N$\mathrm{\acute{e}}$el order. Even at lowest temperatures, these peaks are far broader than resolution limits, resulting in significant diffuse scattering away from the Brillouin zone centers. As temperature is raised, the Lorentzian heights decrease and widths increase, until distinguishable features are lost by T = 25 K. These observations are consistent with previously published results by Zaharko \textit{et al.}\cite{zaharko11}, and essentially as expected from the observation of Lorentzian scattering in our own triple-axis measurements\cite{macdougall11}. In Fig.~\ref{fig:DNS_sim_diffuse}(d), we demonstrate this explicitly by simulating the expected diffuse scattering pattern from assuming isotropic Lorentzian scattering at N$\mathrm{\acute{e}}$el Bragg positions, with heights and widths extracted from Fig.~\ref{fig:HB1a}. All major features of the diffuse scattering data are accounted for by this basic model, with no need to assume a second contribution to the scattering. Notably, among these reproducible features are the apparent `streaks' of scattering connecting peak positions (Figs.~\ref{fig:DNS_sim_diffuse} (e) and (f)) which arise from overlapping Lorentzian tails from different Brillouin zones.

Our second set of polarized measurements employed the full XYZ polarization analysis over a smaller region of reciprocal space that encompasses (0 0 2) and (1 1 1) Bragg positions. From the resultant data, we extracted correlations associated with spins within (i.e. the `y'-direction) and perpendicular to (`z'-direction) the [H H L] scattering plane at two temperatures, and plot the results in Fig.~\ref{fig:DNS_mag_diffuse}. In addition to confirming the magnetic origin of diffuse correlations, our results show that spin directions are largely isotropic. Following the analysis of Schweika\cite{schweika10}, we also compared spin-flip polarization in the $x$-direction to scattering taken with polarization reversed to search for signatures of chirality in this direction, but failed to see anything of note. Magnetic correlations are qualitatively similar above and below the temperature T = $T^*$, consistent with our claim that diffuse scattering is largely the result of Lorentzian-like correlations, which we have shown exist at all temperatures.

\section{Discussion}

The data presented in this article provide a clear picture of how magnetic correlations develop in the spinel aluminates. Lower than expected ordered moment notwithstanding, the observed correlations in $\mathrm{MnAl_2O_4}$ adhere extremely closely to expectations for a classical antiferromagnet. Bragg peaks are resolution limited at low temperatures and have the symmetry of the predicted N$\mathrm{\acute{e}}$el state for this material. Both energies and intensities of collective excitations in the ordered state are well explained by classical spin-wave theory, and the $J_1$ and $J_2$ implied by fits reproduce the measured ordering temperature. At all temperatures, but especially near $T_N$, the system also exhibits significant short-range Lorentzian-like correlations about Bragg positions, which can be completely understood within the context of classical critical fluctuations associated with the second order phase transition.

\begin{figure}[ht]
\begin{center}
\includegraphics[width=\columnwidth]{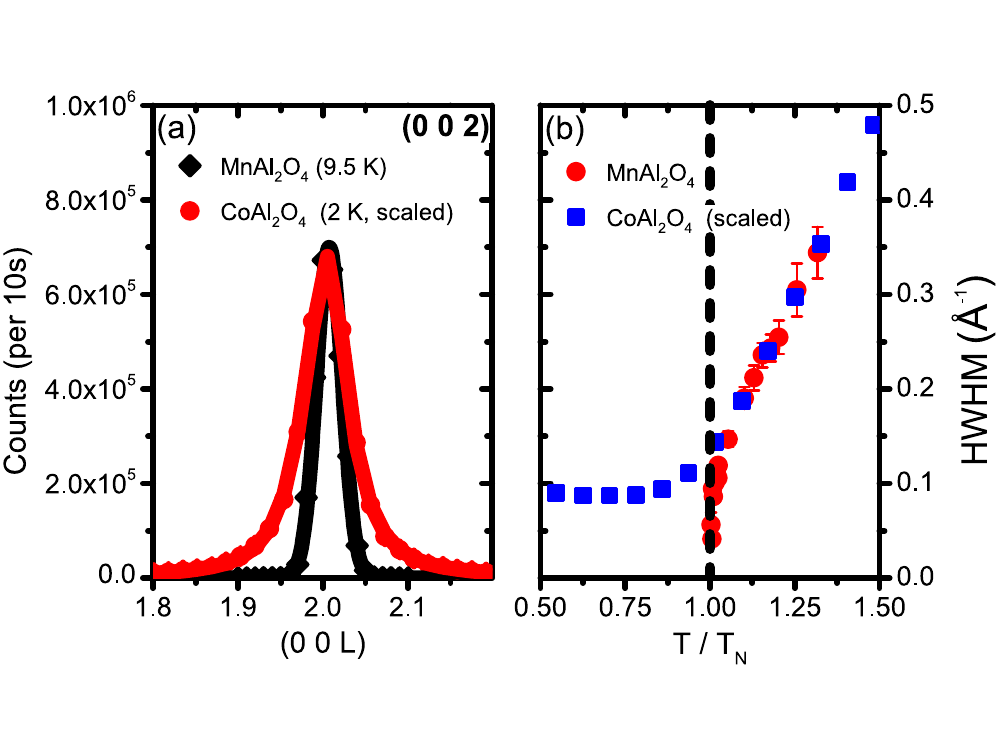}
\caption{Direct comparison of antiferromagnetic correlations in $\mathrm{MnAl_2O_4}$ and $\mathrm{CoAl_2O_4}$ single crystals. \textbf{(a)}: Plots of scattering intensity across the (0 0 2) magnetic Bragg peaks at base temperature for the two materials, and \textbf{(b)}: plots showing the temperature dependence of the antiferromagnetic correlation length, derived from the width of observed Lorentzian scattering peaks. In panel (a), data for $\mathrm{CoAl_2O_4}$ has been scaled along the intensity axis to allow for a more direct comparison of the two materials. In panel (b), data for $\mathrm{CoAl_2O_4}$ has been scaled along both width and temperature axes.}\label{fig:compare}
\end{center}
\end{figure}

Given the strong similarities between the two systems, it is very difficult then to associate the Lorentzian-like correlations in $\mathrm{CoAl_2O_4}$ as a signature of anything but critical correlations heading towards a similar ordered state. Indeed, as shown in Fig.~\ref{fig:compare}, the variation of the correlation widths (inverse of correlation length) with temperature is identical between the two samples when well above $T^*$. The distinction comes at lower temperatures, approaching and below the putative ordering transitions. In $\mathrm{MnAl_2O_4}$, the magnetic correlation length diverges as the N$\mathrm{\acute{e}}$el temperature is approached from above, as expected for a second order transition; below this temperature, resolution-limited Gaussian Bragg peaks emerge. The antiferromagnetic correlation length in $\mathrm{CoAl_2O_4}$  plateaus instead of diverging, indicative of a freezing transition; this is despite the lower density of defects in this compound when compared to $\mathrm{MnAl_2O_4}$. At temperatures T $<$ T$^*$, the magnetic scattering peaks remain broader than resolution and have an anisotropic Lorentzian-squared character. This empirical description is successful in describing the data, regardless of sample or instrument resolution, and remains unchanged to temperatures as low as T = 0.25 K. The inelastic spectrum is dominated by collective spinwave excitations, indicative of an ordered state. As discussed thoroughly above, we believe the best explanation of these collective results is the freezing out of domain wall motion below a first-order phase transition, whose thermodynamic signatures are presumably suppressed due to the close proximity to the known Lifshitz point. This is an effect of increased $\frac{J_2}{J_1}$ in $\mathrm{CoAl_2O_4}$, however the conditions which  led to this unusual state were sufficiently general that we expect similar phenomenology might arise in any number of frustrated spin systems with first-order phase transitions or unconventional glass phases at low temperature.

The largest remaining question then is whether this system displays any definite signature of SSL correlations coexisting with the confirmed N$\mathrm{\acute{e}}$el ordered state. Insomuch as the predicted signature of SSL order is a build-up of incommensurate diffuse correlations tracing out a well-defined manifold in reciprocal space, the firm answer to this question is no: in neither this nor any previous study of $\mathrm{CoAl_2O_4}$ has such a state been observed. One interesting feature in the diffuse scattering patterns of $\mathrm{CoAl_2O_4}$, reported previously \cite{zaharko11} and confirmed in Figs.~\ref{fig:DNS_sim_diffuse} and Fig.~\ref{fig:DNS_mag_diffuse}, is the existence of apparent `streaks' of scattering along crystallographic $<$1 1 1$>$ directions when connecting the antiferromagnetic Bragg positions (but not otherwise). It has previously been suggested that these streaks are a signature of co-existing SSL correlations\cite{zaharko11}, and Monte Carlo simulations argue that this scattering eventually evolves into the predicted spiral surface when $J_2$ is increased\cite{zaharko14}. We caution against such an exotic interpretation, however. We have pointed out that the overlap of Lorentzian tails between neighboring Brillouin zones would also create apparent streaks of scattering, and explicitly demonstrate in Figs~\ref{fig:DNS_sim_diffuse} (e) and (f) that both the qualitative shape and magnitude of these features along line scans similar to Ref.~\onlinecite{zaharko11} can be reproduced within this simplest of pictures. We expect more careful modeling which includes slight anisotropic (Lorentzian-squared) spin correlations, non-spherical magnetic form factors and finite instrument resolution width would only improve agreement with experiment. If simply the result of overlapping Lorentzian tails, this implies that the diffuse streaks are yet another signature of the known short-range antiferromagnetic correlations, and further explains why they evolve with the same temperature dependence- opposite the expected behavior for a competing state. In fact, similar diffuse streaks might be expected between Bragg centers due to \textit{any} mechanism which destabilizes long-range order, regardless of the associated local picture, and would only become more distinct if a `line Lorentzian' or comparably sophisticated scattering model was employed to properly account for microscopic defect structures\cite{savici07}.

\section{Conclusions}

In conclusion, we have performed a series of neutron scattering measurements which shed light on the nature of magnetic correlations in two comparable aluminate spinel systems, $\mathrm{MnAl_2O_4}$ and $\mathrm{CoAl_2O_4}$. Results on $\mathrm{MnAl_2O_4}$ are unequivocal, and paint this material as a classical antiferromagnet whose behavior is in-line with expectations for a weakly frustrated diamond lattice system. This includes observations of a second-order transition to a collinear N$\mathrm{\acute{e}}$el ground state with resolution limited Bragg peaks, Lorentzian-like critical correlations, and collective spin-wave excitations with predictive dispersions and scattering intensities. This is in direct contrast to measurements of $\mathrm{CoAl_2O_4}$ which, despite the material having a lower level of disorder, reveal that long-range antiferromagnetic correlations are arrested at a freezing transition $T^* = 6.5$ K. Below this temperature, we show that the scattering function develops a second, anisotropic Lorentzian-squared contribution- an empirical analysis that we confirm is independent of sample, independent of instrument resolution and persists to temperatures as low at T = 0.25 K. We argue that this is a signature of a frozen coarsening state, where the movement of domain walls is inhibited below a first-order phase transition, in turn preventing the formation of true long-range order. The stark contrast between $\mathrm{CoAl_2O_4}$ and $\mathrm{MnAl_2O_4}$ emphasizes that the novel physics in $\mathrm{CoAl_2O_4}$ is driven primarily by frustration associated with an increased $\frac{J_2}{J_1}$ ratio.

\section{Acknowledgements}

Research at the High Flux Isotope Reactor was sponsored by the U. S. Department of Energy, Office of Basic Energy Sciences, Scientific User Facilities Division. G.J.M is further supported by the National Science Foundation, under grant number DMR-1455264-CAR.  The work in NHMFL (HDZ) was supported by NSF-DMR-1157490 and the State of Florida. DGM acknowledges support from the U.S. Department of Energy, Office of Science, Basic Energy Sciences, Materials Sciences and Engineering Division.

%\bibliography{../spinels_fullauthors}

\end{document}